# Continuously Shaping Orbital Angular Momentum with an Analog Optical Vortex Transmitter

**Kun Huang[1*], Hong Liu[2*], Muhammad Q. Mehmood[1], Shengtao Mei[1], Aaron Danner[1], Jinghua Teng[2†], and Cheng-Wei Qiu[1†]**

[1]Department of Electrical and Computer Engineering, National University of Singapore, 4 Engineering Drive 3, Singapore 117576, Singapore
[2]Institute of Materials Research and Engineering, Agency for Science, Technology and Research, 3 Research Link, Singapore 117602, Singapore
[3]NUS Graduate School for Integrative Sciences and Engineering, National University of Singapore, 28 Medical Drive, Singapore 117456

* These authors contributed equally to this work
† Correspondence and requests for materials should be addressed to C.W. Q. (email: eleqc@nus.edu.sg) or J. H. T. (email: jh-teng@imre.a-star.edu.sg).

Dynamic generation of obitial angular momentum (OAM) of light has enabled complex manipulation of micro-particles, high-dimension quantum entanglement and optical communication. We report an analog vortex transmitter made of one bilaterally symmetric grating and an aperture, emitting optical vortices with the average OAM value continuously variant in the entire rational range. Benefiting from linearly-varying transverse dislocation along its axis of symmetry, this diffractive transmitter possesses extra degree of freedom in engineering broadband optical vortices meanwhile preserving a novel spiniform phase with equally spaced singularities. It unlimitedly increases the average OAM of light by embracing more singularities, which is significantly different from that for Laguerre-Gaussian (LG) and Bessel vortex beams. Realizing analog generation of OAM in a single device, this technique can be potentially extended to other frequencies and applied to a wide spectrum of developments on quantum physics, aperiodic photonics and optical manipulation.

*Introduction* The orbital angular momentum (OAM) of light has been manifested via helical wavefront with phase singularity as a distinct identity of optical vortices. Since its discovery in 1992 [1], the OAM of light has excited a surge of interest because it brings a new degree of freedom and unbounded quantum states for light beams. The current technology trend has been perceived to direct from fundamental investigations towards probing viabilities of harnessing the OAM of light for applications, such as optical trapping and manipulation [2-8], photon entanglement [9-12], astronomy [13] and microscopy [14-15], remote sensing and detection [16-17], optical communications [18-19] and even integrated photonics [20-27]. The fast-moving exploitation on such diverse areas has pushed for further development on OAM generation technology.

Hitherto, the device for OAM generation is primarily based on discretization although one can theoretically tune the OAM continuously by changing the topological charge for LG and Bessel beams [28-30] or tailoring the ellipticity of Ince-Gaussian modes [31]. A typical optical vortex has a helical phase $e^{i\ell\varphi}$ (where $\ell$ and $\varphi$ are topological charge and angular coordinate, respectively) [1], which is angularly continuous for integer $\ell$ while discontinuous with a phase step along $\varphi=0$ for fractional $\ell$. This leads to a similar influence on their generators, e.g., fork gratings having smoothly variant fringes for integer $\ell$ and a cut of fringes with discontinuity along $\varphi=0$ for fractional $\ell$ [32-33]. This distinction between integer and fractional $\ell$ fundamentally makes it impossible to transfer integer OAMs to fractional ones in a single device, resulting in poor reconfigurability—different OAM states must be individually addressed by separate devices. These discrete generators include the traditional fork gratings by computer-generated hologram (CGH) [34-35], spiral phase plates [36], q-plate [37], and the miniaturized devices such as metasurface vortex plate [20-22], nanostructured spiral plate [23-25], integrated vortex emitter [26] and topological nanoslits [27]. The digital devices such as spatial light modulator (SLM) [38] and digital micro-mirror device (DMD) [39] have been employed for CGH to acquire multiple OAMs in practice. However, their intrinsic pixel resolution limit leads to spatial phase jump and also accounts for inaccuracy of fractional OAM, which can be referred to Supplementary Materials (SM).

On the other hand, continuously tunable generation of OAM has gained much attention for its great potentials in applications like quantum entanglement [40] and optical micro-manipulation [41]. Remarkably, attempts have been made to acquire tunable OAM via indirect manners such as weighted superposition of two cross-polarized beams [41], interference of two vortices [42], interaction of spin-orbital angular momentum via internal conical diffraction [43], and optical geometric transformation using two cascaded phase elements [40, 44-46]. Although they can offer a new degree of control for the OAM of light, these approaches are intrinsically accompanied with either poor coherence, very limited tunable range, or complicated transformation with indispensable optical correction after a long-distance propagation. Apparently, a concrete analog methodology, i.e., continuous generation of arbitrary OAM, is still pending for a feasible solution.

Here we report an analog transmitter comprising of bilaterally symmetric gratings with an aperture to produce unique vortex beams carrying rational OAM without any theoretical limit. Differing from other vortex beams (e.g. LG and Bessel beams) that change their OAM by its topological charge of helical phase around one singularity, our vortex beam is imprinted with a spiniform wavefront with unlimited phase singularities equally located in a line and tunes its average OAM via the number of involved singularities. The approach realizes analog generation of OAM in the rational range by embracing these phase singularities smoothly through the aperture. This practical approach might revolutionize the methodology of manipulating the OAM states of light from discretization to continuity basis, and would benefit across quantum physics and optics.

***Design Principle*** Traditionally, light with a plane wavefront can acquire its OAM by successively passing through $\ell$ concentric and vertically located SPPs each of which has the topological charge (TC) of 1, and finally the light possesses a helical wavefront having one single singularity with TC of $\ell$ [32]. As a result, its carried OAM can be controlled by the strength (i.e., $\ell$) of the phase singularity [28-30]. Similarly, light could also obtain OAM by passing through a series of transversely located SPPs, having a wavefront with spatially separated singularities, as shown in Fig. 1(a). One can increase its OAM by involving more SPPs, leading to more phase singularities in its wavefront. Hence, we infer that, if a phase profile with regularly distributed (e.g., periodic) singularities is encoded into one single vortex generator, optical vortices would be emitted smoothly by employing a gradually varying aperture, resulting in the continuous generation of OAM.

To realize it, we propose a bilaterally symmetric grating with an aperture as a vortex transmitter whose working principle is schematically shown in Fig. 1(b). With its *y*-axis being the line of symmetry, this two-dimensional transmitter consists of two periodic binary gratings with a tilting angle $\gamma$. A circular aperture is placed above the gratings tangential to *x* axis at the origin point $O$ and its diameter $d_q$ can vary along *y* axis without breaking its bilateral symmetry. Upon a normally incident plane wave, the transmission function of the grating can be expressed as

$$T(x,y) = 1/2 + \sum_{n=1}^{\infty} \text{sinc}\left(\frac{2n-1}{2}\right) \cos\{(2n-1) \cdot [\kappa_x x - \text{sgn}(x)\beta y]\}, \quad (1)$$

where $\text{sinc}(x)=\sin(\pi x)/(\pi x)$, the diffraction order $n$ is a positive integer, $\kappa_x$ is a constant determining the diffraction angle, $\beta$ stands for a constant phase gradient along $y$ direction, $\text{sgn}(x)$ refers to the sign function of variable $x$ and is mathematically responsible for bilateral symmetry of structure. Detailed in the SM, the grating parameters such as the period $\Lambda=2\pi/(\kappa_x^2+\beta^2)^{1/2}$ and the inclination angle $\gamma=\tan^{-1}(\beta/\kappa_x)$, can be derived.

Light from its 1st-order diffraction (i.e., $n=1$) possesses a linearly $y$-dependent phase item:

$$\chi(x,y) = e^{\text{sgn}(x)i\beta y}, \quad (2)$$

where $\text{sgn}(x)$ accounts for the opposite phase variation tendency at any two symmetric positions ($\pm x_0$, $y_0$) along the $y$-direction: $\chi$ increases at positive values of $x$ while decreases at negative values of $x$. In order to unveil its phase singularities, we show the phase profile after a low-pass filter (see SM) in Fig. 1(c), removing the phase jump along $y$-axis of symmetry. Due to its linear dependence on $y$, phase difference between both sides is periodically formed along the interface, leading to phase singularities equally spaced by a spatial interval $\tau$. Within one phase cycle of $2\pi$, the amplitude of phase jump reaches its maximum of $\pi$ at phase singularity for twice, which means that the phase difference by spanning a distance of $\tau$ along $y$ is $\beta\tau=\pi$, i.e. $\tau=[\Lambda\cdot(1+\tan^2\gamma)^{1/2}]/(2\tan\gamma)$ after some algebraic formulations.

Acting as a regulator for continuous generation of optical vortices, the aperture smoothly changes its diameter along $y$ axis of symmetry to precisely control the linear output of phase with singularities. To quantify this output, we introduce a dimensionless parameter: singularity strength $q \equiv d_q/\tau$. Because the aperture size $d_q$ can be tuned smoothly, $q$ sequentially varies its integral and fractional values so as to realize continuous generation of optical vortices by a single transmitter. We plot the phase along the circumference of aperture for different $q$ in Fig. 1(d), showing a net phase change of $2\pi[q]$, where $[q]$ denotes the round-up of $q$ and equals the number of phase singularities encircled in the aperture. As expected, our results in Fig. 1(e) reveal that the generated vortex beam occupies an average OAM of $Q\hbar$ ($\hbar$ is the reduced Planck constant) of a photon with

$$Q = 0.7q - \sin(\pi q/2) \cdot \text{sinc}(\pi q/2)/2, \quad (3)$$

which will be discussed in details later.

**Theoretical and Experimental Verification** For a normally incident plane wave with electric field $U(x,y,z=0)$ (where $x$, $y$, $z$ are the Cartesian coordinates), the electric field of diffracted light can be theoretically described by angular spectrum method [47]

$$u(x,y,z) = \iint_{\infty} \hat{A}(\kappa_x, \kappa_y) \cdot e^{i(k_x x + k_y y + k_z z)} dk_x dk_y \quad (4)$$

where

$$\hat{A}(k_x, k_y) = \frac{1}{4\pi^2} \iint_{\infty} U(x,y,0) \cdot g(x,y) \cdot e^{-i(k_x x + k_y y)} dxdy \quad (5)$$

where $\hat{A}(k_x, k_y)$ is the two-dimensional Fourier transform of the electric field, $g(x,y)=T(x,y)\cdot\text{circ}(r_d/d_q)$, $r_d^2=x^2+(y-d_q/2)^2$, $\text{circ}$ is the circular function describing the aperture and $k_x$, $k_y$ and $k_z$ denote respective components of wave vector $k$ of a plane wave.

In the experiments, we apply the following specifications to sample fabrications: $\Lambda$=1μm, $\gamma$=tan$^{-1}$(1/240) and correspondingly $\tau$=120μm. This transmitter was patterned on 100-nm thick chromium film deposited on a quartz substrate via electron beam lithography and dry etching process. To achieve high-fidelity experimental results, the apertures were directly fabricated on transmitter, leaving the individual samples with different $q$. Two groups of specimens meant for integral and fractional $q$ were fabricated to exemplify the analog generation concept of rational OAM. The integer group comprised of 4 samples with aperture diameters $d_q$ varying from 120μm to 480μm at a step of 120μm, corresponding to integer $q$ =1~4, respectively; the fraction group comprised of 6 samples with their aperture diameters $d_q$ varying from 132μm to 180μm, which correspond to $q$=1.1~1.5, respectively. The scanning electron microscopy (SEM) images of fabricated samples are provided in SM.

Figure 2 shows the simulation and experimental results of intege group ($q$=1~4) at 532 nm wavelength. Under the assumption of uniform illumination with $U(x,y,0)$=1 for all cases, the simulated intensity and phase profiles of light from the 1$^{st}$ order diffraction at far field (Fraunhofer region) are shown in Fig. 2(a). At $q$=1, elliptical transverse profile is formed and it splits into a two-lobe shape from $q$=2 onwards. As $q$ rises, the central dark region expands itself to accommodate more phase singularities, moving the two lobes farther apart. Meanwhile, these two lobes shrink due to gradually weak diffraction when the aperture continues to enlarge [47]. The simulated intensity profiles are well reconstructed by measurements as shown in Fig. 2(b).

Optical wavefront of the produced beam were experimentally revealed by interference with a co-propagating reference Gausssian beam via Mach-Zehnder interferometer, as shown in Fig. 2(c). Under interferences, dislocated fringes for plane-wave case and spiral arms for spherical-wave case have been respectively revealed in the interferograms as expected, manifesting the existence of phase singularities. The respective topological charge is quantifiable through the number of dislocated fringes for plane-wave case or the arms number for spherical-wave case. Note that the simulated interference patterns (see SM) agree very well with measured results. Figure 2(d) plots the corresponding phase profiles retrieved from experimental results of plane-wave interference through Fourier transformation [48]. The retrieval was validated by simulation, exhibiting nearly identical phase distributions as plotted in Fig. 2(e). The phase shift accumulated along a closed circle is 2π·$q$, which quantifies the respective integer topological charge of $q$.

Similarly, the fractional group ($q$=1.1~1.5) has been examined under the same conditions and their results are shown in Fig. 3. As $q$ goes up, the predicted transverse intensity profile in Fig. 3(a) evolves from an ellipse to H-shape and its bottom half tends to enclose a dark core, which is a manifestation of phase singularity. It agrees with the measured intensity result depicted in Fig. 3(b), including the experimental interference patterns with a plane wave. It can be observed that dislocated fringes become enhanced as the increment of $q$, which is attributed to the fact that the neighboring singularity gets dominated gradually. The experimentally retrieved phase profiles (Fig. 3(c)) agree well with the simulated results (Fig. 3(a)) on the singularity distribution and phase variation. Beyond this, an animation about continuous generation of these optical vortices with varying $q$ can be found in SM.

The irregular intensity and phase profiles of the proposed vortex beam indicate that it has no well-defined OAM per photon like LG and Bessel beams. To build up the connection between our vortex beam and LG beam, we decompose the spiniform phase in Eq. (2) in term of angular-dependence helical phase

$$\chi(x,y) = \lim_{M\to\infty} \sum_{m=-M}^{M} C_m e^{-im\varphi}, \qquad (6)$$

where the coefficient $C_m = \sum_{k=0}^{\lfloor \frac{M-|m|}{2} \rfloor} \left[ \frac{\text{sgn}(x)\beta r}{2} \right]^{2k+|m|} \frac{(-1)^{k+|m|\cdot\frac{1-\text{sgn}(m)}{2}}}{k!(k+|m|)!}$ and $\lfloor a \rfloor$ is the round-down function denoting the integer part of $a$. Eq. (6) implies that our proposed vortex beam with spiniform phase can be taken as a weighted superposition of LG beam and possesses the average OAM of a photon, which is similar with Ince-Gaussian modes [31].

To further quantify its analog effect, the average OAM carried by this paraxial optical vortex beam has been investigated by both theoretical prediction and experimental measurement, as shown in Fig. 1(e). By using the electric field $u(x, y, z=\infty)$ at far field in Cartesian coordinates (see Appendix) [1], its average OAM (denoted by $Q$ in units of $\hbar$) of a photon is evaluated through Eq. (7) and finally correlated as a one-to-one function of $q$ via Eq. (3) which fits simulated results. Due to the coupling of *sine* and *sinc* functions, it shows non-linearity within the interval of [0, +2], beyond which quasi-linearity governs the relation between $Q$ and $q$ in the rational range. This is distinct from the pure non-linear relationship possessed by LG and Bessel beams [28-30]. Thus, the average OAM carried by this optical vortex beam has been validated to be continuously addressable in rational states without any theoretical limit.

***Discussion*** Propagation stability of the vortex beam, an important factor for complex optical transformation in applications such as optical manipulation [2-8] and photon entanglement [9], is characterized with a fractional case of $q$=1.4, for example. Depicted in Fig. 4(a) are the experimental intensity profiles at selected z-cut planes along propagation direction. Initialized from z=5cm (Fraunhofer distance), the vortex beam propagates with a well-preserved transverse intensity up to a far distance of z=25cm, exhibiting a high stability against propagation.

Moreover, broadband operation of this transmitter has been theoretically investigated, showing that its transmitted light only has the diffraction order of 0 and ±1 for the normally incident visible light ($\lambda$=400nm~800nm), see SM. Using a super-continuum laser, we measured their diffraction efficiency $\eta_j = I_j / \sum_j I_j$ where $j$=0, ±1 and $I_j$ is the intensity of $j^{\text{th}}$ order diffraction. The measured broadband efficiency $\eta_{\pm 1}$ is around 23% with a tiny deviation from the theoretical value 22.4% without wavelength dependence predicted by Eq. (1), as plotted in Fig. 4(b). However, diffraction angle $\alpha$ is wavelength-dependent and proved by both theoretical and experimental results. The diffraction angle $\alpha$ monotonously increases with wavelength $\lambda$ while high fidelity of the elliptical transverse profile is well kept, exemplified by the elliptical transverse profiles at various wavelengths with a sample of $q$=1, as shown in Fig. 4(c).

In particular, this transmitter can potentially be miniaturized to shape relativistic electron beam for a matter-wave vortex, and alternatively upsized to generate OAM-carried radio waves for communication purpose. This device can also be utilized for micro- & nano-patterning via laser ablation and photolithography. Arbitrarily maneuvering OAM in rational states and more importantly in a continuous fashion makes it attractive to enrich OAM coding/encoding techniques for remote sensing, detection and optical encryption.

***Conclusion*** In summary, we have rigorously demonstrated the concept of analog OAM generation and the analog OAM transmitter that is able to produce optical vortices carrying theoretically boundless rational OAMs. Constituted by periodical gratings of bilateral symmetry with a tunable aperture, the transmitter is endowed with aperiodicity to uniquely embed scalable amount of singularities into the phase of the diffracted light, fundamentally owing to the linearly varying transverse dislocation to facilitate smooth phase transition. In addition, the mechanism tailoring the OAM states of light by the number of involved phase singularities provides a unique insight for investigating quantum states of light for quantum physics and singular optics. This unprecedented approach bridges the technology gap of digital, discrete and analog shaping of light's orbital angular momentum.

**Appendix**
**Simulation** All the simulations about the intensity and phase profiles are carried out by using angular spectrum method described in Eq. (4). The far-field Fraunhofer diffraction is realized by using a thin Fourier lens with a long focal length of $f$ [47] and the simulated far-field intensity and phase profiles are obtained at its focal plane. Because it is a paraxial vortex beam in our case, we apply the simulated electric field $u(x,y,z=f)$ at the focal plane to evaluate its average OAM ($Q\hbar$) of a photon in unit length by employing the well-known definition [1]

$$Q = \frac{i}{2} \frac{\iint \left[ x\left(u^*\frac{\partial u}{\partial y} - u\frac{\partial u^*}{\partial y}\right) - y\left(u^*\frac{\partial u}{\partial x} - u\frac{\partial u^*}{\partial x}\right) \right] dxdy}{\iint |u|^2 dxdy}, \quad (7)$$

where $u=u(x,y,z=f)$. The definition of Cartesian coordinates is used as our vortex beam exhibits non-axisymmetric intensity and phase profiles. The relationship between $Q$ and $q$ is predicted by using Eq. (7) and shown by the squares in Fig. 1(e). Its fitting curve with a root mean square error (RMSE) of 0.04 has an analytical form of Eq. (3).

**Amplitude and phase retrieval from experimental interference** Retrieval of amplitude and phase from experimental interferences was based on the filtering technique of spatial-frequency spectrum. After a Fourier transformation, the interference patterns of the plane-wave cases had three dominated orders: 0 and ±1 orders, which were corresponding to the frequency information of the reference and vortex beams, respectively. To retrieve the phase and amplitude of light from the +1 order diffraction, we needed to implement an inverse Fourier transformation of the +1 order frequency information by removing the others, which has been detailed in SM. With the retrieved data of phase and amplitude, we pursued the calculation of the average OAM ($Q$) of a photon through Eq. (7) and plotted it as experimental data denoted by green asterisks of Fig. 1(e).

**Authors Contributions**
K. H., H. L., C. Q. and J. T conceived the idea. K. H carried out simulation and the structural design. H. L. and M. Q. M. fabricated the samples. K. H, H. L. and S. M. measured and analysed the experimental results. K. H., H. L., A. D., C. Q. and J. T. prepared the manuscript. C. Q. and J. T. supervised the overall work. All the authors analysed the data and discussed the results.

**Acknowledgement**
We thank Prof. Miles Padgett and Prof. E. Brasselet for valuable discussions in improving the manuscript. We also thank Dr. Yuxuan Ren for his instructive suggestions in experimental measurement. K. H and C. Q thank the support from CRP project (No.: R-263-000-A86-281) of Singapore NRF. The work is partially supported by the Institute of Materials Research and Engineering and the Agency for Science, Technology and Research (A*STAR) under Grant 1021740172.

**Figure captions**

**FIG. 1 (color online). Concept and mechanism of the analog optical vortex transmitter shaping rational OAM.** **(a)** Light obtaining a helical wavefront with spatially separated phase singularities (black dots) by passing through four transversely located SPPs. Two neighbouring SPPs have a phase difference of $\pi$ for getting a smooth phase profile. One singularity in phase profile means that light passes through one SPP, leading to their number identical to SPPs number. **(b)** Sketch of the optical vortex transmitter composed of two inclining (inclination angle of $\gamma$) gratings with a period of $\Lambda$ at both sides and a circular aperture (just above the gratings) with varying diameter ($d_q$), geometrically playing as an excircle (red dashed circles) tangent to x-axis at a reference point $O$. The ±1 order diffraction beam with an angle α between it and 0-order diffraction has a helical phase. **(c)** Phase profile encoded into the vortex transmitter. $\tau$ denotes the spatial distance between two neighbouring phase singularities. Upon varying the circular aperture (white dashed circle), the diffractive beam is regulated with continuously increased phase contours to acquire more singularities. It eventually leads to a singularity strength $q$ of rational value, i.e. arbitrarily switching between any integer and fraction. $\varphi$ is the angle coordinate of circular aperture and increases anticlockwise from $\varphi=0$ (negative *y* axis) to $2\pi$. **(d)** Phase along the circumference (dashed circle in (d)) of circular aperture for its corresponding $q$. The phase at $\varphi > \pi$ is unwrapped by adding $2\pi$. The curves denote the phase value for different $q$s that are distinguished by their curve colours, showing a smoothly variant $q$. **(e)** The average OAM ($Q\hbar$) of a photon as a function of singularity strength $q$. The fitting curve (solid red line) of simulated results (black square boxes) exhibits a root mean square error of 0.04 while experimental results are denoted by greenish asterisks. Inset: Zoom-in of data between $q=1$ and $q=1.5$.

**FIG. 2 (color online). Optical vortices with integer $q$.** **(a)** Simulated transverse intensity and phase profiles for $q=1$, 2, 3 & 4, respectively. **(b)** Measured intensity profiles at the far field. **(c)** Experimental interference patterns with plane (left) and spherical (right) waves. **(d)** Phase profiles retrieved from experimental interference patterns with plane wave. $\Phi$ denotes the angular coordinate. **(e)** Quantitative comparison of azimuthal phase shift ($\Delta P \equiv P(\Phi)-P(\Phi=0)$, where $P$ is the phase of this vortex beam) between experiment (curves) and simulation (square boxes). Data is obtained along the black dashed circles as shown in (d).

**FIG. 3 (color online). Optical vortices with fraction $q$.** **(a)** Simulated intensity and phase profiles for $q=1.1$, 1.2, 1.25, 1.33, 1.4 & 1.5, respectively. **(b)** Measured intensity profiles (left) for different $q$ and their corresponding interference patterns (right) with plane wave. **(c)** Phase profiles reconstructed from experimental interference with plane wave.

**FIG. 4 (color online). Propagation stability of vortex beam and broadband behaviour of the transmitter.** **(a**) Normalize experimental transverse intensity profiles at selective z-cut planes (z=5cm, 10cm, 15cm, 20cm & 25cm) for verification of propagation stability exemplified by $q=1.4$. To facilitate a clear visualization, each image is post-processed with transparency for low (below 0.2) intensity. The inset (*z*=0) schematically shows the binary mask (not to scale). **(b)** Measured (dots) and theoretical (line) diffraction efficiency of this transmitter in visible range. **(c)** Measured (dots) and theoretical (curve) diffraction angle of optical vortex for broadband light. Insets: (left) experimental intensity profiles for *q*=1 at different wavelengths.

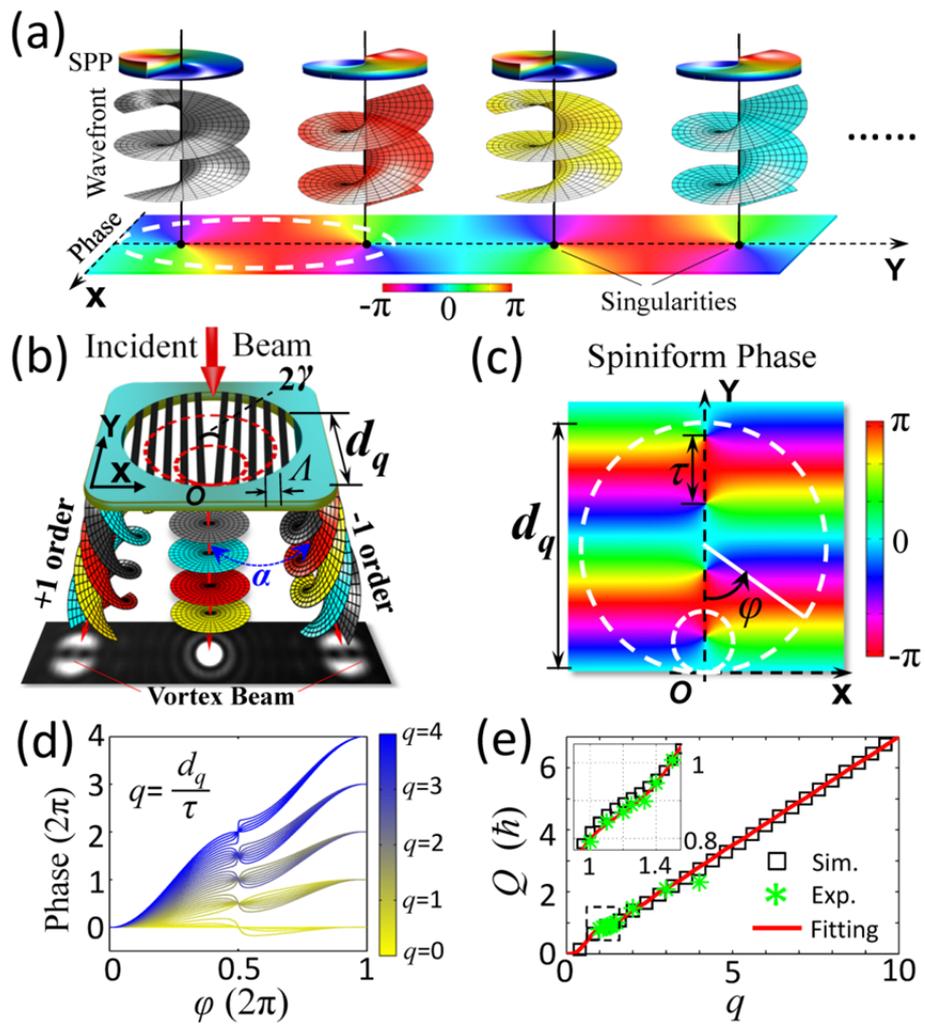

Figure 1

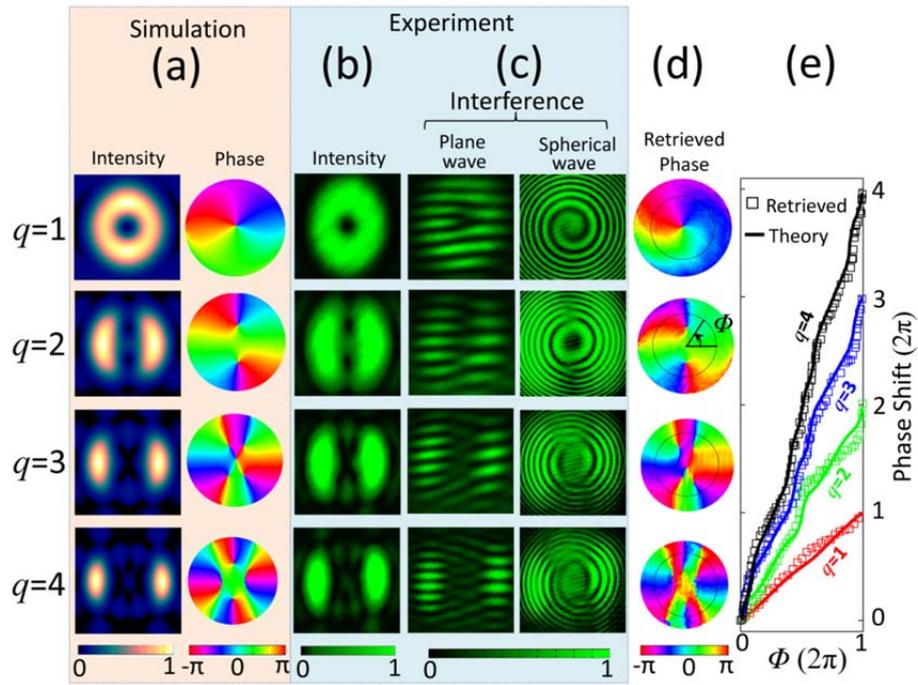

Figure 2

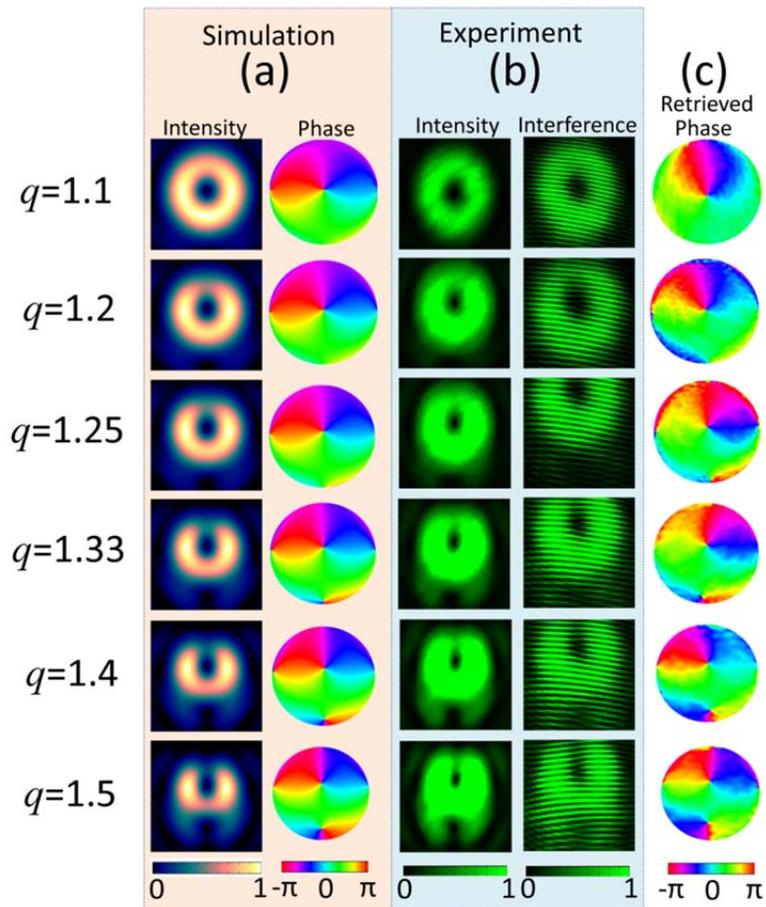

Figure 3

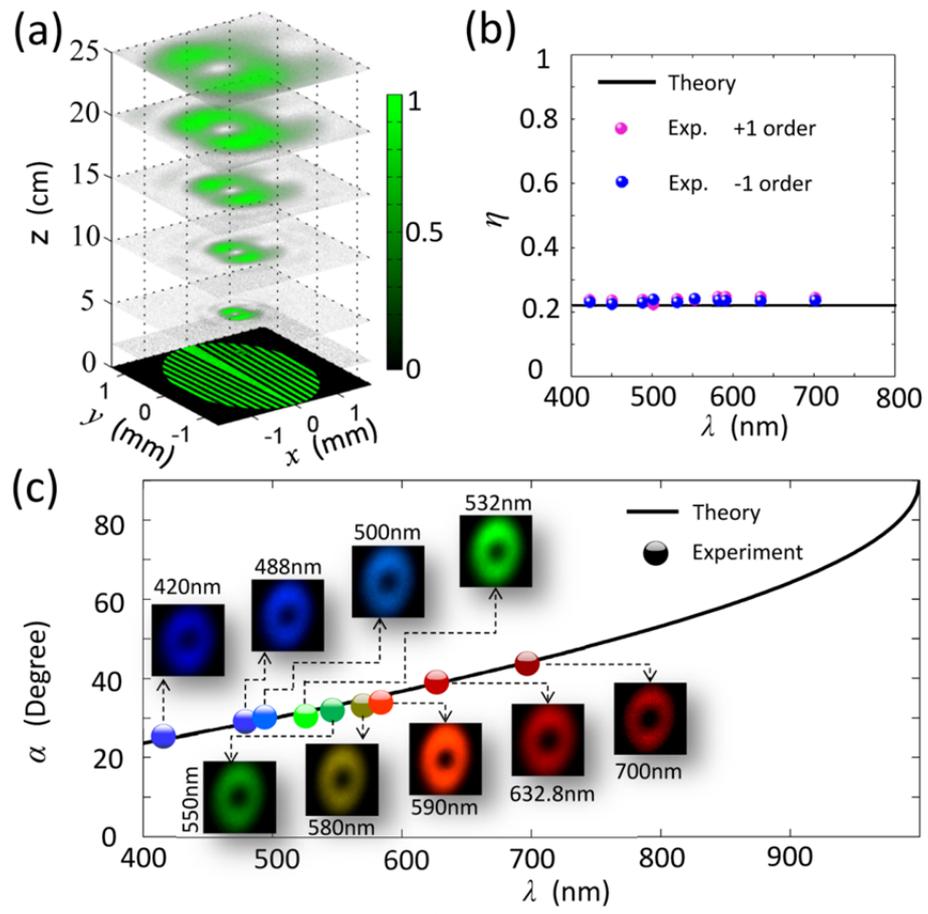

Figure 4

# Supplementary Materials

## 1. Bilateral Symmetrical Structure

Structure mathematically described by Eq. (1) in the main text has a bilaterally symmetric pattern as shown in Fig. S1(a). We can get the basic information (i.e. grating period $\Lambda$ and inclination angle $\gamma$) of this vortex transmitter in Eq. (1), where the item relative with the spatial coordinates of $x$ and $y$ is

$$f(x,y) = \kappa_x x - \mathrm{sgn}(x)\beta y, \tag{S1}$$

which is responsible for the structural parameters (i.e. grating period $\Lambda$ and inclination angle $\gamma$) of transmitter. To unveil this, Eq. (S1) can be further modified as

$$f(x,y) = \sqrt{\kappa_x^2 + \beta^2}\left[\cos(\gamma)x - \mathrm{sgn}(x)\sin(\gamma)y\right], \tag{S2}$$

where $\cos(\gamma) = \kappa_x/\sqrt{\kappa_x^2 + \beta^2}$ and $\sin(\gamma) = \beta/\sqrt{\kappa_x^2 + \beta^2}$, leading to $\tan\gamma = \beta/\kappa_x$. Considering that Eq. (1) has the form of $\cos[f(x,y)]$, we can directly get the grating period $\Lambda = 2\pi/\sqrt{\kappa_x^2 + \beta^2}$. Thus, we have the constant $\beta$ and $\kappa$ (in terms of $\Lambda$ and $\gamma$)

$$\beta = \frac{2\pi \tan\gamma}{\Lambda\sqrt{1 + \tan^2\gamma}}, \tag{S3}$$

$$\kappa_x = \frac{2\pi}{\Lambda\sqrt{1 + \tan^2\gamma}}. \tag{S4}$$

This means that constant $\beta$ depends on the grating period $\Lambda$ and its inclination angle $\gamma$. Once the grating is fixed, one can get the parameters $\beta$ and $\kappa_x$, which respectively determines the distance between two phase singularities and its broadband behaviour that will be discussed in details later.

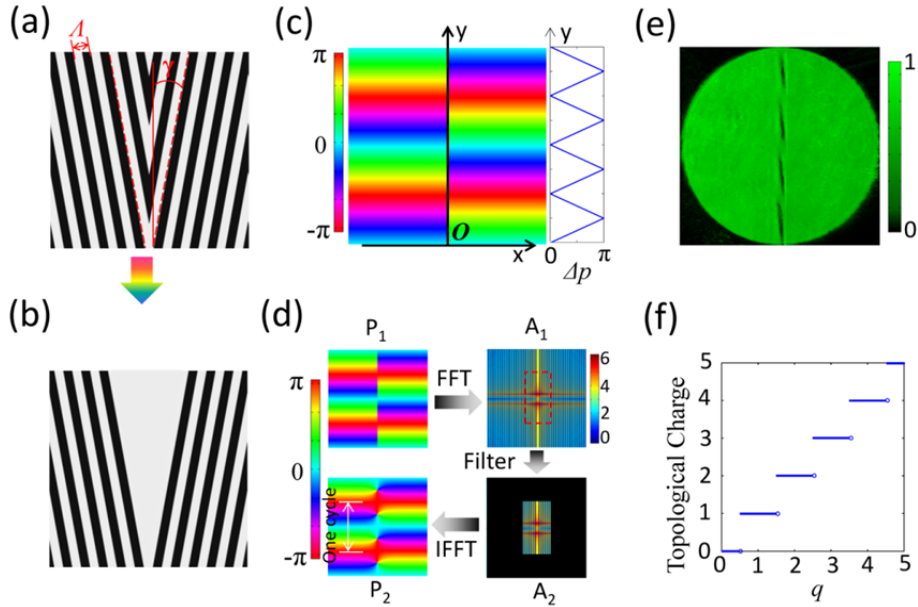

**Figure S1| Vortex transmitter and its encoded phase.** (**a**) Sketch of vortex transmitter directly described by Eq. (1) in the main text. The white and black bars represent transparent and opaque areas of the transmitter. The triangles in between two dashed red lines are removed to form the proposed transmitter, as sketched in (b). (**c**) The encoded phase profile $P(x,y)$ directly described by Eq. (2) in the main text. A clear phase jump from positive $x$ to negative $x$ happens at the y axis. Inset: Phase difference $\Delta P$ along $y$ between negative x and positive x. $\Delta P = P(x,y)-P(-x,y)$. (**d**) The low-pass filtering operation for unveiling the phase singularities. $P_1=\arg<\chi(x,y)>$ is the phase profile in (c). $A_1=\log_{10}|FFT(P_1)|$ denotes the frequency spectrum and the logarithm operation is employed for a better show of frequency. $A_2$ is the low-passed frequency spectrum obtained by selecting the spectrum in the dashed red rectangle and padding zero (black) in other regions. $P_2=IFFT(A_2)$ shows the obvious phase singularities. (**e**) Experimental intensity profile captured by an objective lens (10X, 0.3 NA) slightly above the exit plane of transmitter with $q$=4. The four dark regions mean four phase singularities. (**f**) Relationship between topological charge of phase enclosed in a circular aperture and its corresponding $q$. It shows only integer topological charge is obtained.

In order to decrease the technical requirement in fabrication, we remove all the structures between two dashed lines in Fig. S1(a) and make it transparent in our practical vortex transmitter shown in Fig. S1(b). Thus, the acute vertexes along y axis will not appear in our transmitter, which will significantly decrease the fabrication difficulty when the structure is scaled down to a small size. Interestingly, the generated vortex beam at the far field is nearly the same as that generated by structure in Fig. S1(a). This originate from two aspects: 1) the far-field performance of our transmitter is dominated by two bilaterally symmetrical gratings that bring the inversed phase, leading to a dislocation; 2) light at the far field is usually not influenced by the slight defect at the near field. It can refer to the simulated and measured vortex beams (Fig. 2 in main text) that behave consistently with each other as expected.

The encoded phase profile described by Eq. (2) of main text is shown in Fig. S1(c), which has an obvious phase jump from negative $x$ to positive $x$. The phase difference between both sides is plotted in the inset of Fig. S1(c), showing a periodic change with the largest value of π. The phase singularities exist at the positions with phase difference of π. For both sides of $y$ axis, the changing speed of phase is determined by the gradient β. According to Fourier optics [S1], this phase jump always implies a high spatial frequency in frequency domain, which cannot be maintained at the propagation of light that can be taken as a low-pass filter in free space. In this case, we implement a low-pass filtering of phase profile of Fig. S1(c) to remove this phase jump and unveil the hidden phase singularities. The operation details are shown in Fig. S1(d). Apparently, the filtered phase has a smooth change around the $y$ axis while keeps the phase unchanged in other regions, leading to a spinform vortex phase with two singularities in one cycle. This implies a simple result between $\tau$ and $β$: $\tau β=π$. Then we have $\tau=[\Lambda \cdot (1+\tan^2\gamma)^{1/2}]/(2\tan\gamma)$, which shows a straightforward relationship between the phase singularities and the structural parameter ($\Lambda$ and $\gamma$) of transmitter.

These enclosed phase singularities can be experimentally manifested by the zero-intensity (dark) regions in intensity profiles of Fig. S1(e), captured slightly above the exit plane of vortex transmitter for the exemplary case $q$=4. Because the phase difference between both sides is periodically formed along $y$ aixs as shown by the inset of Fig. S1(c), the dark regions around phase singularities in the captured intensity profiles extends vertically and forms several dark slits with equal interval. Due to Guoy phase shift [S2], these dark slits are a little tilting because the intensity is captured slightly above the exit plane.

We show the relationship between the topological charge and $q$ in Fig. S1(f). Its topological charge is valued at integer [$q$], which is distinguished from Laguerre-Gaussian and Bessel beams with a helical phase ($e^{im\varphi}$) that are able to have fractional topological charge. This can be interpreted by the smooth of our spiniform phase, which obeys the definition of topological charge by M. V. Berry. However, with the increment of $q$, the jump of its topological charge between integers will not impose an influence on the smooth generation of its carried OAM that is tightly dependent on both intensity and phase of vortex beam, which has been clearly shown in Fig. 1(e) and Eq. (3) of main text.

### 3. Fabrication of vortex transmitter

In order to show the high-fidelity experimental result, we fabricated the apertures directly on the vortex transmitter with their vertical distance of zero, so that the error from light diffraction between them can be eliminated. But, this will not affect the concept of analog generation because the double grating in the transmitter is fixed and only the aperture is changed. In fact, when the grating period is large (e.g. up to 20μm), the whole size of transmitter will increase correspondingly up to several millimetres, allowing the traditional iris aperture to realize the smooth output of optical vortices.

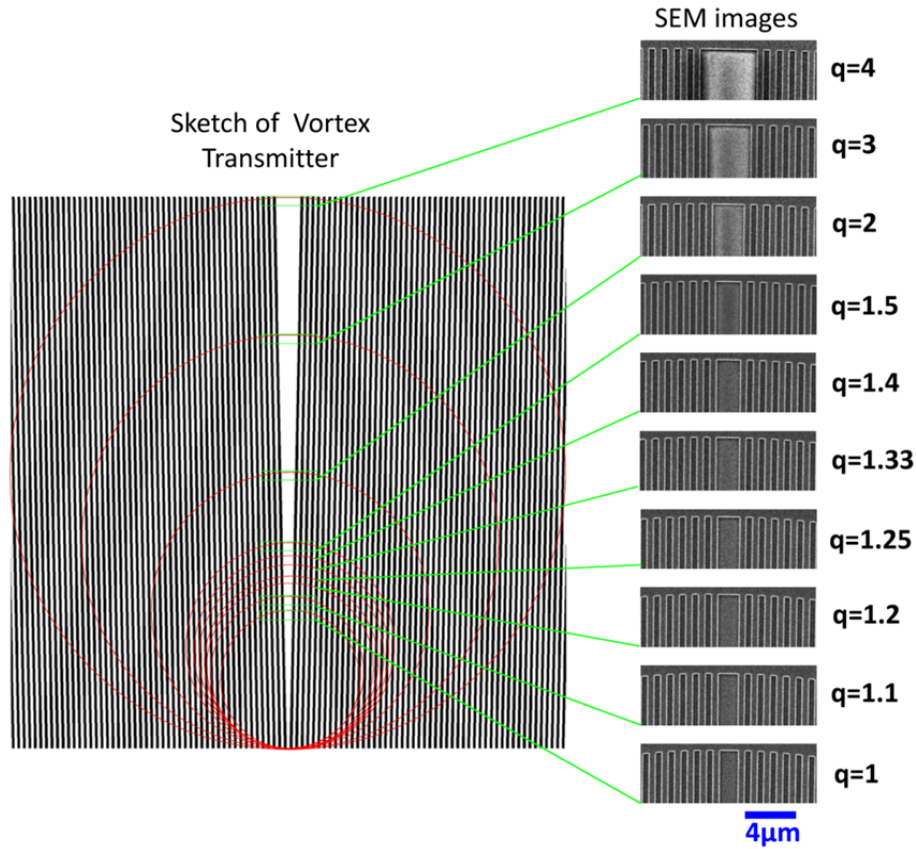

**Figure S2| Fabricated samples of vortex transmitter.** The apertures denoted by the red circles are directly fabricated on the transmitter, leaving individual samples whose SEM images with 3000X magnification are provided in right volume. Correspondingly, the integer group has individual samples of q=2 ($d_2$=240 μm), q=3 ($d_3$=360 μm) and q=4 ($d_4$=480 μm) while the fractional group consists of q=1.1 ($d_{1.1}$=132 μm), q=1.2 ($d_{1.2}$=144 μm), q=1.25 ($d_{1.25}$=150 μm), q=1.33 ($d_{1.33}$=159.6 μm), q=1.4 ($d_{1.4}$=168 μm) and q=1.5 ($d_{1.5}$=180 μm), respectively. The scale bar is available for all SEM images.

The SEM images of fabricated samples are shown in Fig. S2. To better exhibit the details, we only provide the top parts of SEM images for all the fabricated samples. Because the aperture increases, their top parts widen gradually, as shown by their images. The fabrication processes are shown below:

The binary mask was fabricated on 100 nm thick chromium (Cr) film which was deposited by electron beam evaporator (Denton Vacuum, Explorer) at room temperature under a vacuum pressure of 5 x $10^{-7}$ Torr on top of a quartz substrate. ZEP-520A resist mixed with anisole (1:1 ratio) was spin-coated over Cr at a speed of 5000 rpm and then soft-baked at 180°C on a hotplate for a period of 120s. After that, it underwent electron beam lithography (ELS-7000, Elionix) to expose the specimen at a dosage of 240 μC/cm$^2$, beam current of 50 pA and acceleration voltage of 100 kV followed by development in o-Xylene for a period of 30s. The specimen was then subjected to a dry etching process (Nanoquest, Intlvac) under conditions of beam voltage/current of 300 V/110 mA, acceleration voltage/current of 100 V/6 mA, power of 180 W and vacuum pressure of about 2.54 x $10^{-4}$ Torr for approximately 7 minutes. After that, the residual resist was removed by soaking the specimen in Microposit™ remover 1165 (Shipley) for 12 hours followed by oxygen plasma etching (Oxford RIE) under a power of 100 W and O$_2$ flow of 80 sccm for approximately 2 minutes.

4. **Experiment for characterizing the vortex beam**

In order to reveal the phase structure of the generated vortex beam by our transmitter, we have built up an experimental setup for measuring its interference with a plane and spherical wave via Mach-Zehnder interferometer, as shown in Fig. S3. Light from a laser is coupled into a mono-mode fibre and then collimated by an objective lens $L_1$ with its magnification of 5×. The collimated light with its diameter of about 11mm is divided into two beams by a 50-50 beam splitter (BS). One beam as a main beam is used to illuminate the vortex generator. Its transmitted light will form a slightly divergent optical vortex beam, which is collimated by a lens $L_2$ for a better interference pattern, especially for the case of the reference beam with a plane wave. The collimated vortex beam is reflected by a mirror $M_1$ for its normal incidence on another beam-splitter $BS_2$. The vortex beam passing through $BS_2$ is taken as the main beam for interfering with the reference beam.

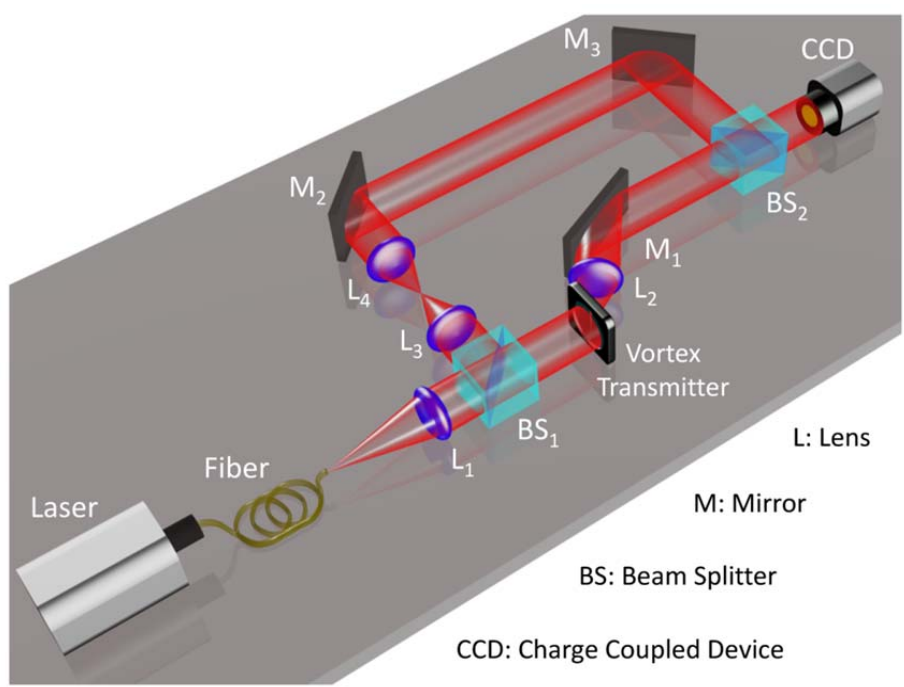

**Figure S3|  Mach-Zehnder interferometer for measuring the phase profile.**

The other beam splitted from $BS_1$ is used as the reference beam. This beam is manipulated by two lenses ($L_3$ and $L_4$) with the same focal length of 50mm. In order to tune the wavefront (plane wave or spherical wave) of reference beam in experiment, we fix the lens $L_3$ and move the lens $L_4$ along the propagating direction of light. When these two lenses are confocal at a same position, light after $L_4$ is a collimated beam as a plane wave. Then, we move the lens $L_4$ forward (backward) along the propagating direction of light, resulting that a focused (divergent) spherical wave is generate. After the reflection precisely tuned by two mirrors ($M_2$ and $M_3$), the reference beam is normally incident on the 50-50 $BS_2$. One part of reference beam from $BS_2$ co-propagates with the main beam in a slight tilting angle so as to observe the interference pattern by a charge-coupled device (CCD).

If the reference beam is tuned for a plane wave, the interference pattern captured by the CCD is the dislocated fringes as shown in Fig. S4 (a), which indicates a good agreement with simulated results in Fig. S4(b). The direction of fringes in the interference pattern can be changed by slightly tuning the mirrors ($M_2$ and $M_3$). Also, the width of fringes can be tuned by changing the angle between the reference and main beam with the help of the mirror M$_3$. The wide fringe means a small angle.

If the reference beam has a wavefront of a spherical wave, its interference pattern has the spiral pattern with different arms of $q$ for various vortex beams, which is shown in Fig. S4(c). The good consistency between simulation and experiment confirms the validation of our method.

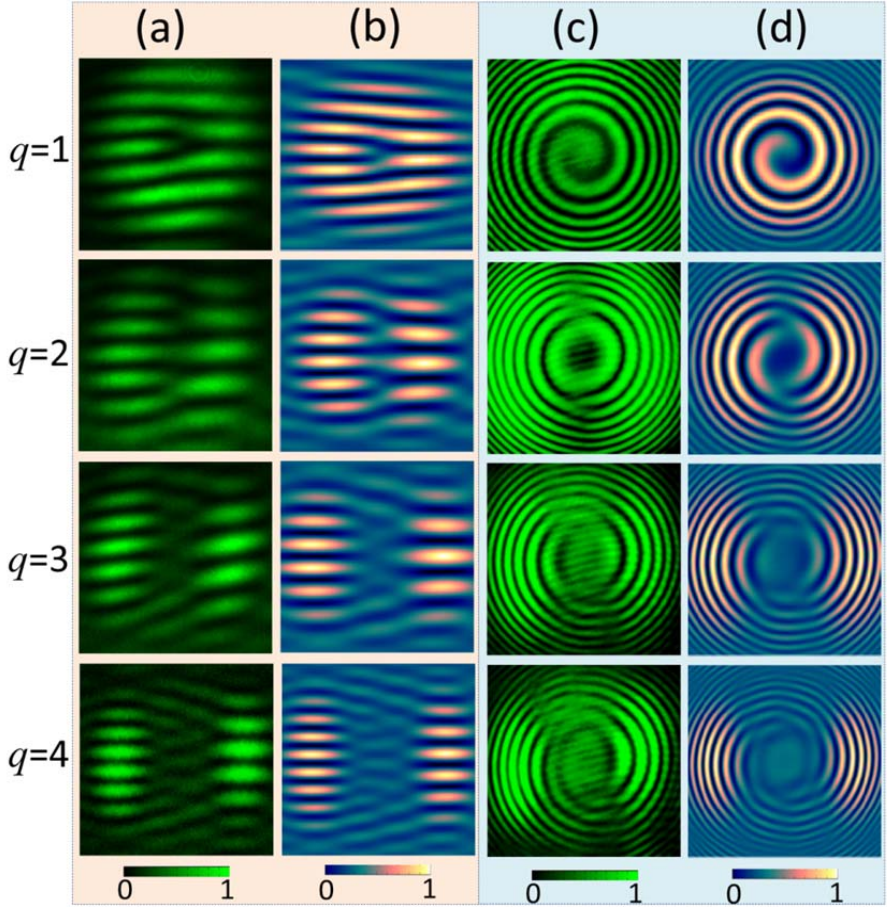

**Figure S4| Interference patterns.** Experimental (a) and simulated (b) interference patterns between our vortex beam and a plane wave. Experimental (c) and simulated (d) interference patterns between our vortex beam and a spherical wave.

## 5. Phase retrieval from experimental interference

To obtain the phase from experimental interference, we use the signal processing technique with the cropping of frequency spectrum on the basis of Fourier transformation. Ignoring its principle theory (one can refer to Ref. [S3]) for this technique, we only introduce several key steps as shown in Fig. S5(a):

Step 1: Fourier transformation of experimental interference. Fourier transformation is realized by fast Fourier transformation (FFT) that shows a good approximation. In addition, experimental interference pattern with dense fringes is preferred for a better retrieval because this vortex beam for this case is encoded into high-frequency information that is apparently separated from the low-frequency, leading to an easier operation in selecting and cropping its frequency spectrum. In the diagram for frequency spectrum of Fig. S5(a), the top and bottom white rectangles respectively denote the frequency information of $\pm q$ vortex beam while the green rectangle stand for that of reference beam. In practice, we only need to retrieve the information of one beam such as $+q$ vortex.

Step 2: Selecting the useful frequency spectrum (denoted by white dashed rectangle), removing the other frequency spectrum, shifting the useful frequency spectrum to the centre and generating the new frequency spectrum for final retrieval. The detailed operations are sketched in Fig. S5.

Step 3: Inverse Fourier transformation of cropped frequency spectrum. Correspondingly, the inverse FFT is employed to carry out the final retrieval from the cropped frequency spectrum. Thus, we can get the retrieved amplitude and phase profiles from the experimental interference, as shown in Fig. S5(a).

The above retrieval steps are only suitable for the fringed pattern interfered with a plane wave because the frequency domain of vortex beam is separated from the low frequency carried by the reference beam. For the spherical-wave cases, it is difficult by using this technique because the frequency information of vortex beam and reference beam overlays with each other.

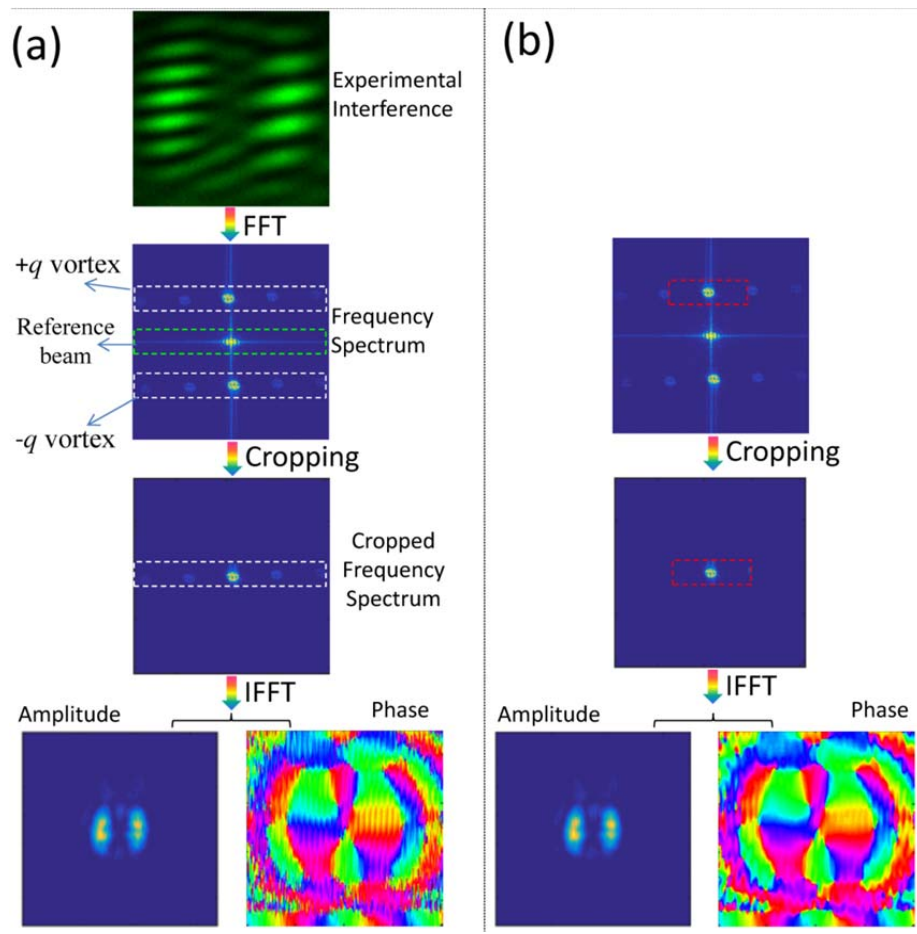

**Figure S5| Phase retrieval from experimental interference. (a)** Steps in carrying out the phase retrieval. It contains three steps: 1) Fourier transformation of experimental interference; 2) Cropping the useful frequency spectrum and shifting it to the centre; 3) Inverse Fourier transformation of cropped frequency spectrum. The case for $q$=3 vortex beam is taken as an example. **(b)** Phase retrieved from a small-sized cropped frequency spectrum without high frequency. This allows us to get a smoother phase pattern.

Another important issue is that, the size of selected frequency area has an obvious influence on the final retrieved results, especially for the smoothness of phase profile. To show this, we select a small sized frequency area for retrieval, as depicted in Fig. S5(b), keeping other operations the same with those in Fig. S5(a). Its smoothness of retrieved phase is better than that of the large-sized case while the amplitude has no obvious change, which shows that the phase of light is more sensitive than amplitude.

6. **Calculating orbital angular momentum carried by our vortex beam**

Considering the fact that our optical vortex beam is paraxial, we use the theory by Allen et al. to calculate its OAM [S4]. Because of the non-axisymmetry of our vortex beam, we evaluate its OAM by using the version in Cartesian coordinates.

For a vector potential *A* polarized along *x*,

$$\mathbf{A} = \mathbf{e_x} \cdot u(x, y, z)\exp(ikz), \tag{S5}$$

the paraxial approximation, after use of the Lorentz gauge for convenience [S5], yields:

$$\mathbf{B} = \mu_0 \mathbf{H} = \nabla \times \mathbf{A} = \begin{bmatrix} \mathbf{e}_x & \mathbf{e}_y & \mathbf{e}_z \\ \dfrac{\partial}{\partial x} & \dfrac{\partial}{\partial y} & \dfrac{\partial}{\partial z} \\ u \cdot e^{ikz} & 0 & 0 \end{bmatrix} = \left(\dfrac{\partial u}{\partial z}e^{ikz} + ikue^{ikz}\right)\mathbf{e}_y - \dfrac{\partial u}{\partial y}e^{ikz}\mathbf{e}_z$$

$$\approx ike^{ikz}\left(u\mathbf{e}_y + \dfrac{i}{k}\dfrac{\partial u}{\partial y}\mathbf{e}_z\right) \tag{S6}$$

$$\mathbf{E} = -i\omega\mathbf{A} - \nabla\cdot\left(\dfrac{i}{\omega\varepsilon_0\mu_0}\nabla\cdot\mathbf{A}\right) = -i\omega u e^{ikz}\mathbf{e}_x - \nabla\cdot\left(\dfrac{ic^2}{\omega}\dfrac{\partial u}{\partial x}e^{ikz}\right)$$

$$= -i\omega u e^{ikz}\mathbf{e}_x - \dfrac{ic^2}{\omega}\left[\dfrac{\partial^2 u}{\partial x^2}e^{ikz}\mathbf{e}_x + \dfrac{\partial^2 u}{\partial x\partial y}e^{ikz}\mathbf{e}_y + \left(\dfrac{\partial^2 u}{\partial x\partial z}e^{ikz} + ik\dfrac{\partial u}{\partial x}e^{ikz}\right)\mathbf{e}_z\right]$$

$$= \omega\left\{-iue^{ikz}\mathbf{e}_x - \dfrac{ic^2}{\omega^2}\left[\dfrac{\partial^2 u}{\partial x^2}e^{ikz}\mathbf{e}_x + \dfrac{\partial^2 u}{\partial x\partial y}e^{ikz}\mathbf{e}_y + \left(\dfrac{\partial^2 u}{\partial x\partial z}e^{ikz} + ik\dfrac{\partial u}{\partial x}e^{ikz}\right)\mathbf{e}_z\right]\right\} \tag{S7}$$

$$\approx \omega\left\{-iue^{ikz}\mathbf{e}_x - \dfrac{i}{k^2}\left[ik\dfrac{\partial u}{\partial x}e^{ikz}\mathbf{e}_z\right]\right\} = -i\omega\left(u\mathbf{e}_x + \dfrac{i}{k}\dfrac{\partial u}{\partial x}\mathbf{e}_z\right)e^{ikz}$$

Then, we have some basic formulas about the cross product of **E** and **B**:

$$\mathbf{E}^* \times \mathbf{B} = \begin{bmatrix} \mathbf{e}_x & \mathbf{e}_y & \mathbf{e}_z \\ i\omega u^* & 0 & \dfrac{\omega}{k}\dfrac{\partial u^*}{\partial x} \\ 0 & iku & -\dfrac{\partial u}{\partial y} \end{bmatrix} = \left(-i\omega u\dfrac{\partial u^*}{\partial x}\mathbf{e}_x + i\omega u^*\dfrac{\partial u}{\partial y}\mathbf{e}_y + k\omega|u|^2\mathbf{e}_z\right), \tag{S8}$$

$$\mathbf{E} \times \mathbf{B}^* = \begin{bmatrix} \mathbf{e}_x & \mathbf{e}_y & \mathbf{e}_z \\ -i\omega u & 0 & \dfrac{\omega}{k}\dfrac{\partial u}{\partial x} \\ 0 & -iku^* & -\dfrac{\partial u^*}{\partial y} \end{bmatrix} = \left(i\omega u^*\dfrac{\partial u}{\partial x}\mathbf{e}_x - \omega u\dfrac{\partial u^*}{\partial y}\mathbf{e}_y + k\omega|u|^2\mathbf{e}_z\right), \tag{S9}$$

Using Eqs. (S8) and (S9), we get the linear momentum density as follows:

$$\varepsilon_0 \langle \mathbf{E} \times \mathbf{B} \rangle = \frac{\varepsilon_0}{2} \left[ (\mathbf{E} \times \mathbf{B}^*) + (\mathbf{E}^* \times \mathbf{B}) \right]$$

$$= \frac{\varepsilon_0}{2} \left( i\omega u^* \frac{\partial u}{\partial x} \mathbf{e}_x - \omega u \frac{\partial u^*}{\partial y} \mathbf{e}_y + k\omega |u|^2 \mathbf{e}_z - i\omega u \frac{\partial u^*}{\partial x} \mathbf{e}_x + i\omega u^* \frac{\partial u}{\partial y} \mathbf{e}_y + k\omega |u|^2 \mathbf{e}_z \right), \quad \text{(S10)}$$

$$= \frac{i\varepsilon_0 \omega}{2} \left[ \left( u^* \frac{\partial u}{\partial x} - u \frac{\partial u^*}{\partial x} \right) \mathbf{e}_x + \left( u^* \frac{\partial u}{\partial y} - u \frac{\partial u^*}{\partial y} \right) \mathbf{e}_y \right] + \varepsilon_0 k\omega |u|^2 \mathbf{e}_z$$

$$\varepsilon_0 (\mathbf{r} \times \langle \mathbf{E} \times \mathbf{B} \rangle) = [x, y, z] \times \left\{ \frac{i\varepsilon_0 \omega}{2} \left[ \left( u^* \frac{\partial u}{\partial x} - u \frac{\partial u^*}{\partial x} \right) \mathbf{e}_x + \left( u^* \frac{\partial u}{\partial y} - u \frac{\partial u^*}{\partial y} \right) \mathbf{e}_y \right] + \varepsilon_0 k\omega |u|^2 \mathbf{e}_z \right\}$$

$$= \left[ y\varepsilon_0 k\omega |u|^2 - z \frac{i\varepsilon_0 \omega}{2} \left( u^* \frac{\partial u}{\partial y} - u \frac{\partial u^*}{\partial y} \right) \right] \mathbf{e}_x + \left[ -x\varepsilon_0 k\omega |u|^2 + z \frac{i\varepsilon_0 \omega}{2} \left( u^* \frac{\partial u}{\partial x} - u \frac{\partial u^*}{\partial x} \right) \right] \mathbf{e}_y ,$$

$$+ \frac{i\varepsilon_0 \omega}{2} \left[ x \left( u^* \frac{\partial u}{\partial y} - u \frac{\partial u^*}{\partial y} \right) - y \left( u^* \frac{\partial u}{\partial x} - u \frac{\partial u^*}{\partial x} \right) \right] \mathbf{e}_z$$

(S11)

Therefore, according to Allen's definition, the ratio of angular momentum to energy per unit length of beam can be expressed as

$$\frac{J_z}{W} = \frac{\varepsilon_0 \iint dxdy (\mathbf{r} \times \langle \mathbf{E} \times \mathbf{B} \rangle)_z}{c\varepsilon_0 \iint dxdy (\langle \mathbf{E} \times \mathbf{B} \rangle)_z} = \frac{\iint M_z dxdy}{\iint j_z dxdy}, \quad \text{(S12)}$$

where

$$M_z = \frac{i\varepsilon_0 \omega}{2} \left[ x \left( u^* \frac{\partial u}{\partial y} - u \frac{\partial u^*}{\partial y} \right) - y \left( u^* \frac{\partial u}{\partial x} - u \frac{\partial u^*}{\partial x} \right) \right], \quad \text{(S13)}$$

$$j_z = c\varepsilon_0 k\omega |u|^2, \quad \text{(S14)}$$

Finally, we have the ratio

$$\frac{J_z}{W} = \frac{i}{2\omega} \frac{\iint \left[ x \left( u^* \frac{\partial u}{\partial y} - u \frac{\partial u^*}{\partial y} \right) - y \left( u^* \frac{\partial u}{\partial x} - u \frac{\partial u^*}{\partial x} \right) \right] dxdy}{\iint |u|^2 dxdy}, \quad \text{(S15)}$$

which is usually used to evaluate the average OAM (in unit of $\hbar$) of photon in unit length.

By using Eq. (S15), we can calculate the average OAM carried by our vortex beam once its electric field with amplitude and phase is given. For simulation, the electric field can be easily obtained by employing Eq. (4) in the main text. However, its electric field in experiment can only be retrieved from its experimental interference, which has been described above. All the simulated and experimental results about its average OAM are provided in Fig. 1(f) of main text. In addition, we also give a curve fitting of simulation results with a root mean square error of 0.04 and find that its OAM per photon has an analytical dependence on $q$: $Q=0.7q-\sin(\pi q/2)\cdot\text{sinc}(\pi q/2)/2$, which shows a clear description about its average OAM value and also finally confirms the analog generation of rational OAM by using our proposal.

For Laguerre-Gaussian and Bessel beams having a helical phase of exp($il\varphi$) (where $l$ is the topological charge and $\varphi$ is the angle coordinate), their carried OAMs (in units of $\hbar$) have the form of $L=l\cdot\sin(2l\pi)/(2\pi)$ [S6-S8], showing a non-linearity dependence on its topological charge $l$. For integer $l$, they have a well-defined OAM of $l\hbar$. But for fractional $l$, $\sin(2l\pi)/(2\pi)$ works as a modification item for its carried OAM. This is mainly because the phase exp($il\varphi$) for fractional $l$ can be taken as a weighted super-position of exp($il\varphi$) with integer $l$. In addition, optical vortex beams with fractional $l$ have the non-axisymmetric intensity profiles, which are induced by the discontinuity of helical phase in angular position. As a result, when $l$ varies from an integer to its neighbouring integer, LG and Bessel beams have their intensity profiles changing from axisymmetry, to non-axisymmetry and axisymmetry, showing a regular variation tendency.

However, for our vortex beam, there is no such obvious tendency. For the case $0<q<1$, its phase and intensity have a similar behaviour (from a plane wave for $q=0$ to a vortex beam with a ring intensity for $q=1$) with that of LG and Bessel beam so that the *sine* item of our case dominates this region. Then, when $q$ increases from 1 to 2, its intensity profile changes from a ring shape to two-lobe shape, showing a non-linear variation. This physically explain the non-linear dependence of $Q$ on $q$ in the range from $q=0$ to $q=2$ in Fig. 1(f). For $q>2$, the vortex beam keeps a two-lobe intensity, having a linear variation that is dominated by the *sinc* item. Correspondingly, the item $\sin(\pi q/2)\cdot\text{sinc}(\pi q/2)/2$ tends to be zero for $q>2$, leaving a linear item of $0.7q$.

7. **Broadband behaviour of our vortex transmitter**

We find that our analog vortex transmitter is broadband. Eq. (1) in the main text can be expressed as

$$T(x,y) = \frac{1}{2} + \frac{1}{2}\sum_{n=1}^{\infty}\text{sinc}\left(\frac{2n-1}{2}\right)\exp\{i(2n-1)\cdot[\kappa_x x - \beta y\,\text{sign}(x)]\}$$
$$+ \frac{1}{2}\sum_{n=1}^{\infty}\text{sinc}\left(\frac{2n-1}{2}\right)\exp\{-i(2n-1)\cdot[\kappa_x x - \beta y\,\text{sign}(x)]\}, \quad (S15)$$
$$= A_0 + \sum_{n=1}^{\infty}A_{n+} + \sum_{n=1}^{\infty}A_{n-}$$

where

$$A_0 = 1/2, \quad (S16)$$

$$A_{n+} = \frac{1}{2}\text{sinc}\left(\frac{2n-1}{2}\right)\exp\{i(2n-1)\cdot[\kappa_x x - \beta y\,\text{sign}(x)]\}, \quad (S17)$$

$$A_{n-} = \frac{1}{2}\text{sinc}\left(\frac{2n-1}{2}\right)\exp\{-i(2n-1)\cdot[\kappa_x x - \beta y\,\text{sign}(x)]\}, \quad (S18)$$

which denote the zero-order, positive and negative-order diffraction, respectively. The non-zero order diffraction contains the phase item of $\chi(x,y) = e^{\text{sgn}(x)i\beta y}$, which is independent on the wavelength of light. It means that one can get this vortex beam at the non-zero order diffraction for a broadband light, as sketched in Fig. S6(a). For the non-zero order diffraction, it is a tilting wave with a phase item of $\chi(x, y) = e^{\text{sgn}(x) i \beta y}$. Its tilting angle between this wave and $x$ axis is denoted by $\theta$, as shown in Fig. 6S(a). From the structural information of this transmitter, this tilting wave is determined by the phase item of $\exp[i(2n-1)\kappa_x x]$. On the other hand, this tilting wave can also described by using the phase item of $\exp[i2\pi/\lambda \cos\theta \cdot x]$ from the viewpoint of light propagation in free space. This leads to an equality $(2n-1)\kappa_x = 2\pi/\lambda \cos\theta$. After $\kappa_x$ is substituted by Eq. (S4), we arrive at

$$\cos\theta = (2n-1)\cos(\gamma)\lambda/\Lambda, \qquad (S19)$$

which indicates the broadband behavior of our vortex transmitter, showing that the angle $\theta$ is dependent on the wavelength. Eq. (S19) is a very important result for predicting the broadband behavior that is discussed in more details as follow.

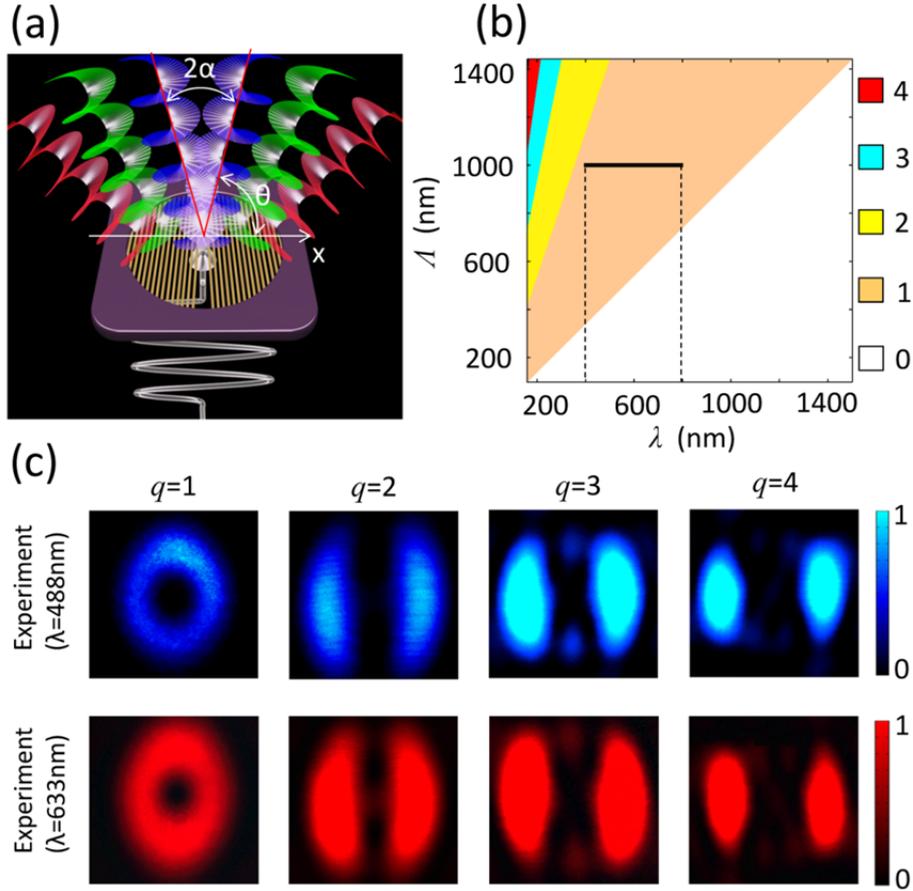

**Figure S6| Broadband behaviour of our vortex transmitter. (a)** Sketch for its broadband behaviour. A white light is incident on our structure, leading to multiple vortex beams with different wavelength. **(b)** Dependence of the largest diffraction orders $n_{\max}$ on incident wavelength ($\lambda$) and grating period ($\Lambda$). **(c)** The measured intensity profiles for different $q$ at the wavelength 488nm and 633nm.

**7.1. Diffraction order**. Physically, angle $\theta$ is valued in the range between $[0, \pi/2]$. From Eq. (S19), one can predict the largest diffraction order $n_{\max} = \lfloor (\Lambda/(\cos\gamma\lambda)+1)/2 \rfloor$, where $\lfloor x \rfloor$ is the round-down function denoting the integer part of $x$. Figure S6(b) shows $n_{\max}$'s dependence on the wavelength $\lambda$ and grating period $\Lambda$ under the assumption of small $\gamma$, leading to $\cos\gamma \approx 1$. For the cases of $n_{\max}=0$, it means that the incident wavelength is larger than the grating period so that there is no high-order diffraction, that is, no vortex beam is generated. This theoretically demonstrates that our vortex transmitter is only valid for light with its wavelength smaller than grating period $\Lambda$.

Especially, our fabricated vortex transmitter with $\Lambda = 1\mu m$ and $\gamma = \tan^{-1}(1/240)$ has only $n_{\max}=1$ in visible range from $\lambda=400$nm to $\lambda=800$nm, as shown in Fig. S6(b).

**7.2. Diffraction angle α**. The angle 2α between ±*n*-order diffraction can be obtained directly by

$$\alpha = \pi/2 - \cos^{-1}\left[(2n-1)\cos(\gamma)\lambda/\Lambda\right] = \sin^{-1}\left[(2n-1)\cos(\gamma)\lambda/\Lambda\right]. \tag{S20}$$

For our fabricated vortex transmitter with $n_{max}=1$, its relationship between angle α and wavelength λ is shown in Fig. 4(c) in the main text and verified experimentally.

**7.3. Diffraction efficiency**. Our vortex transmitter has only 0-order and ±1-order diffraction. Following the definition, we only consider the diffraction efficiency $\eta_j = I_j/\sum_j I_j$ where $j=0, \pm 1$ and $I_j$ is the intensity of $j^{th}$ order diffraction. Theoretically, from Eq. (S15), the intensity $I_j$ ($j=0, \pm 1$) can be obtained by $I_j=|A_j|^2$, where $A_0=0.5$, $A_{\pm 1}=0.5\text{sinc}(1/2)=0.3183$, leading to $\eta_{\pm 1}=22.38\%$, which is wavelength-independent. It is worthy to point out that, the above results are obtained under the assumption that the $I_j$ contains all the intensity. But, in experiment, it is very hard to capture all the intensity profiles due to the limited size (about 1cm in diameter) of the detector. This is the reason for the slight derivation between experiment and simulation as shown in Fig. 4(b) of main text.

In order to further verify its broadband behavior, we measured the intensity profiles at the wavelength of 488nm and 633nm for the cases with $q=1\sim 4$, which is provided in Fig. S6(c). From the experimental result, we can see that the nearly same intensity profiles are obtained by using one single vortex transmitter.

In addition, we also measured their interference patterns with a spherical wave for the case of $q=1$ as shown in Fig. S7. The single-armed spiral patterns are well reconstructed to validate the existence of a helical wavefront with topological charge of 1.

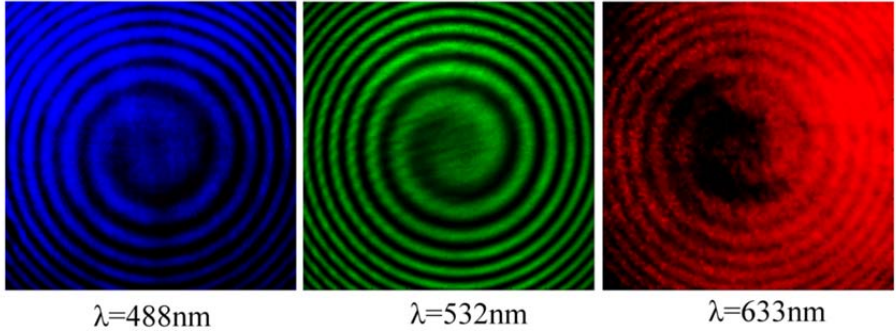

**Figure S7| Measured interference patterns at the wavelength of 488nm, 532nm and 633nm for the case of *q*=1.**

## 8. Inaccurate demonstration of fractional OAM for digital devices (i.e., SLM and DMD)

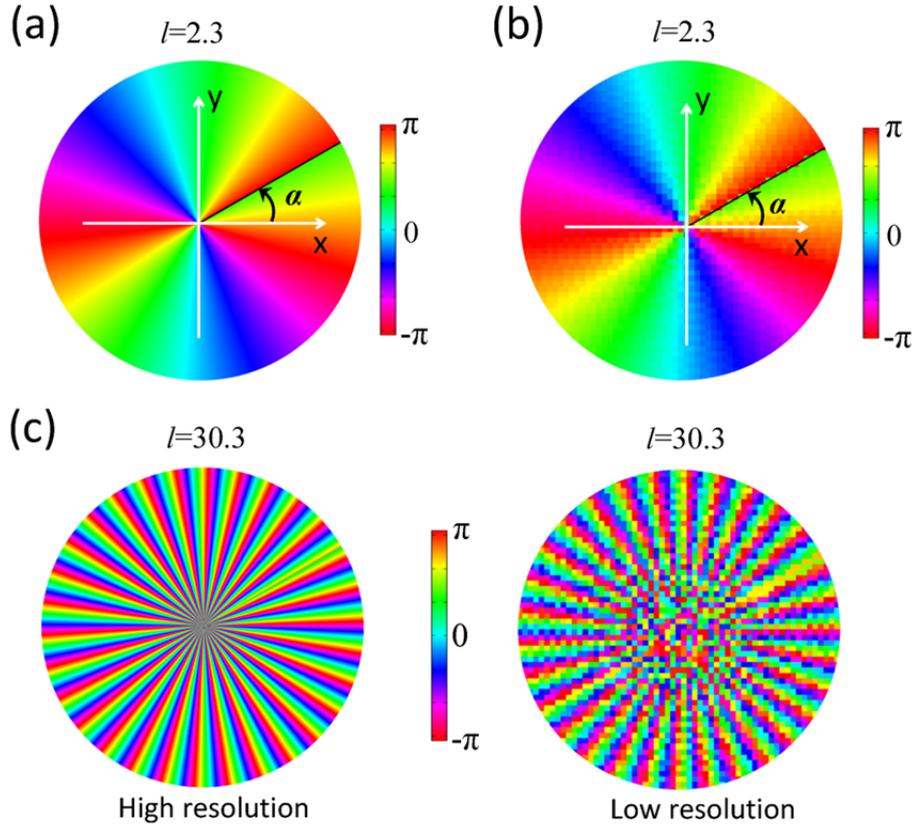

**Figure S8 | Phase profiles of a fractional OAM with high and low resolution.** For low-resolution case, the pixel pitch has the size of 6.4μm*6.4μm, which is the smallest one in industry. (a-b) *l*=2.3 with high and low resolution. (c) *l*=30.3 with high and low resolution.

For a helical phase $e^{i\ell\varphi}$, the phase profile along the angle direction is continuous for integer *l* and continuous for fractional *l* because of phase jump at *φ*=0. More generally, the position of phase jump is located at *φ*=α (black line in Fig. S8(a)), which is an important parameter in implementing quantum entanglement in terms of fractional OAMs [S8]. When one use the digital devices such as spatial light modulator and digital micro-mirror device to demonstrate the helical phase with fractional *l*, their fixed pixel pitches will fail to demonstrate an inaccurate straight line, as shown in Fig. S8(b). Although this inaccurate demonstration of digital devices is usually ignored for small *l*, this inaccuracy will be quite significant for large *l* as shown in Fig. S8 so that it leads to a clear reduction of mode transformation efficiency in quantum entanglement. Therefore, the digital SLM and DMD are not perfect in some special applications such as high-quanta quantum entanglement in terms of orbital angular momentum and optical manipulation.